\documentclass[a4paper,11pt]{article}
\usepackage[skins]{tcolorbox}
\usepackage{mathtools,slashed}
\usepackage[T1]{fontenc}
\usepackage{slashed,verbatim,subfigure}
\usepackage[numbers,sort&compress]{natbib}
\usepackage{amsmath}
\usepackage{mathrsfs}
\usepackage{amsbsy}
\usepackage{amssymb}
\usepackage{appendix}
\usepackage{graphicx}
\usepackage[final]{pdfpages}
\usepackage{dcolumn}
\usepackage{bm}
\usepackage[colorlinks=true,linktocpage=true,
linkcolor=blue,citecolor=blue]{hyperref}
\usepackage[a4paper]{geometry}
\catcode`@11
\def\seceqaa{\@addtoreset{equation}{section}
	\def\theequation{A\arabic{equation}}}
\def\seceqbb{\@addtoreset{equation}{section}
	\def\theequation{B\arabic{equation}}}
\def\seceqcc{\@addtoreset{equation}{section}
	\def\theequation{C\arabic{equation}}}
\def\seceqdd{\@addtoreset{equation}{section}
	\def\theequation{D\arabic{equation}}}
\def\seceqee{\@addtoreset{equation}{section}
	\def\theequation{E\arabic{equation}}}
\catcode`@11
\newcommand{\be}{\begin{eqnarray}}
\newcommand{\ee}{\end{eqnarray}}
\usepackage{authblk}
\begin{document}
\large
\title{\bf Yang-Baxter deformed wedge holography}
\author[,1,2]{Gopal Yadav\footnote{email- gopalyadav@cmi.ac.in, gyadav@ph.iitr.ac.in}\vspace{0.1in}}
\author[,1]{Hemant Rathi\footnote{email- hrathi@ph.iitr.ac.in}}
\affil[1]{Department of Physics, Indian Institute of Technology Roorkee, Roorkee 247667, Uttarakhand, India}
\affil[2]{Chennai Mathematical Institute,
SIPCOT IT Park, Siruseri 603103, India
}

\date{}

\maketitle
\begin{abstract}
In this paper, we construct the wedge holography in the presence of homogeneous Yang-Baxter deformation. We observe that the DGP term is the reason for the existence of non-zero tension of the Karch-Randall branes in Yang-Baxter deformed wedge holography. {The homogeneous Yang-Baxter deformation introduce non-trivial island surfaces inside the black hole horizon whose entanglement entropy is lower than the twice of thermal entropy of the black hole. Therefore, we obtain the Page curve even without the DGP term on the Karch-Randall branes due to the homogeneous Yang-Baxter deformation in the context of wedge holography}. Finally, we compute the the holographic complexity in homogeneous Yang-Baxter deformed $AdS_2$ background. 
\end{abstract}
%\newpage
%\tableofcontents
\section{Introduction}
\label{Introduction}
Duality is a robust proposal in physics. In string theory, the famous duality is the gauge-gravity duality. Gauge-gravity duality is helpful for exploring the strongly coupled gauge theories from the weakly coupled gravitational theories. J. Maldacena proposed the duality between type IIB string theory defined on $AdS_5 \times S^5$ and ${\cal N}=4$ super  Yang-Mills theory \cite{AdS-CFT}. The gauge-gravity duality has been used in many branches of physics, e.g., condensed matter physics, cosmology, quantum chromodynamics (QCD), black hole information paradox, etc. The current paper is focused on the application of an extended version of this duality (wedge holography) to the black hole information paradox.

A very long time ago, Hawking found that black holes do not follow the unitary evolution of quantum mechanics \cite{Hawking,Hawking1}  which give rise to the information paradox. To recover the unitary evolution of black holes, entanglement entropy of Hawking radiation must follow the Page curve \cite{Page}. There has been extensive study on the information paradox using holography, and many proposals have been made (e.g. island proposal, double holography and wedge holography). The island proposal is based on the inclusion of contribution to the entanglement entropy from the interior of black holes at late times \cite{AMMZ}. In early times, islands do not contribute to the entanglement entropy of the Hawking radiation, and one obtains the divergent entanglement entropy. These two phases together (no-island and island) reproduce the Page curve \footnote{See \cite{Swansea,RNBH-HD} and references therein where island proposal has been used.}. In doubly holographic setups, the black hole is living on the end-of-the-world brane, and the external CFT bath acts as the sink to collect the Hawking radiation, see \cite{HD-Page Curve-2,Chou:2021boq,Chou:2023adi} and references therein. Another proposal is based on the wedge holography where the bath is taken as gravitating \cite{WH-i,WH-ii,WH-iii}. In doubly holographic setups and wedge holography, one obtains the Page curve from the entanglement entropies of the extremal surfaces: Hartman-Maldacena \cite{Hartman-Maldacena} and island surfaces. See \citep{massive-gravity,GB-2,GB-3,GB-4,C1,Massless-Gravity,Massless-Gravity-1,Massless-Gravity-2,Massless-Gravity-3,DGP-Swampland,DGP-Swampland-i,Multiverse} for related work on double holography and wedge holography and top-down down double holography in \cite{HD-Page Curve-2} have been obtained for a non-conformal bath (thermal QCD bath) whose gravity dual is ${\cal M}$-theory inclusive of ${\cal O}(R^4)$ corrections \cite{Yadav:2020tyo}.

The purpose of the present paper is to explore the effects of the homogeneous Yang-Baxter deformation in the context of information paradox and holographic complexity. The importance of the YB deformations stems from the fact that they preserve the integrability of the sigma model \cite{Matsumoto:2015jja}\footnote{See also \cite{DR-1} for the study on $\eta$ deformation.}. Recently, the authors in \cite{Kyono:2017jtc} explore the effects of the novel YB deformations in the 2D-dilaton gravity system having a quadratic potential known as Almheiri-Polchinski (AP) model \cite{Almheiri:2014cka}. Interestingly, the authors found that the YB deformed $AdS_2$ background \cite{Kyono:2017jtc} could be a consistent solution of AP model if one deformed the quadratic potential into the hyperbolic function. 

The paper is organized as follows. In section \ref{WH-YBAdS}, we construct wedge holography in the presence of homogeneous Yang-Baxter deformation. Section \ref{IP} compromises of two subsections: \ref{IP-WDGP} and \ref{IP-WDGP-i}. In \ref{IP-WDGP}, we have shown that it is possible to get the Page curve in the presence of homogeneous Yang-Baxter deformation even without DGP term. In subsection \ref{IP-WDGP-i}, we have discussed the DGP term and swampland criteria in homogeneous Yang-Baxter deformed wedge holography. We compute the holographic complexity in homogeneous Yang-Baxter deformation in section \ref{HC} and then summarise our results in section \ref{summary}. There are two appendices \ref{Thermodynamics-YBJT} and \ref{3D-to-2D} on the computation of Hawking temperature and dimensional reduction from 3D to 2D respectively.

\section{Wedge holography in homogeneous Yang-Baxter deformed model}
\label{WH-YBAdS}
\begin{figure}
\begin{center}
\includegraphics[width=0.5\textwidth]{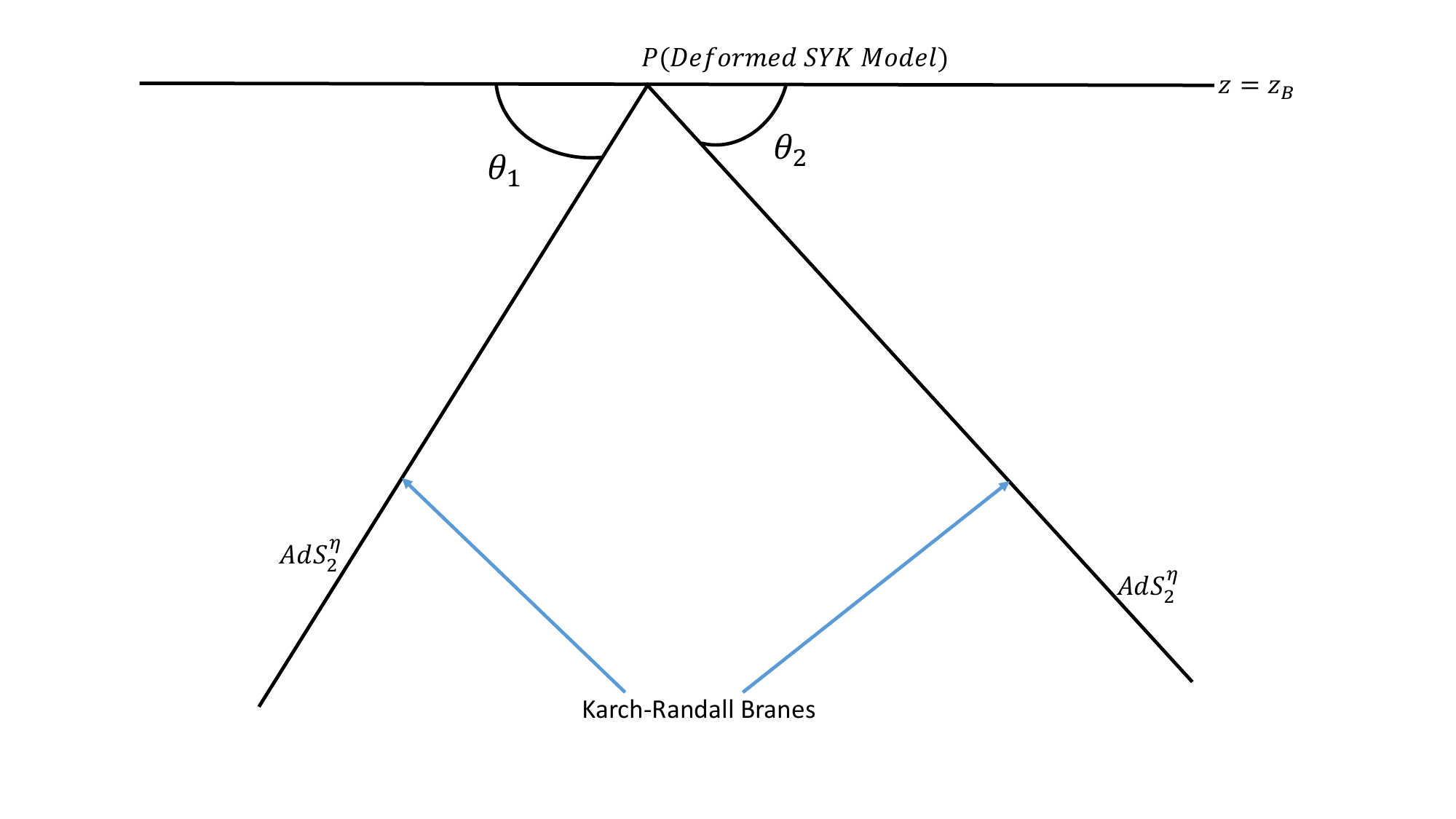}
\end{center}
\caption{Pictorial realization of homogeneous Yang-Baxter deformed wedge holography. $z_B$ is the holographic screen and $P$ is the defect.}
\label{WHS}
\end{figure}

Working action for the wedge holography with $AdS_3$ bulk is written as follows:
\begin{eqnarray}
\label{action-YBD}
S=\frac{1}{16 \pi G_N^{(3)}}\Biggl[\int_M d^3x \sqrt{-g^{(3)}} (R-2 \Lambda) +2\int_{\partial M} d^2 x\sqrt{-h}K+2 \int_{Q_\gamma} d^2x  \sqrt{-h_\gamma}\left({\cal K}_\gamma-T_\gamma\right)\Biggr],
\end{eqnarray}
where the first term of the above equation is Einstein Hilbert term with negative cosmological constant \cite{YB-AdS3}, second term is the Gibbons-Hawking-York boundary term, and third term is defined on the Karch-Randall branes with $\gamma=1,2$ located at constant $\theta$ where Karch-Randall branes have induced metric $h_\gamma$ and tensions $T_\gamma$. On varying the action (\ref{action-YBD}) with respect to metric ($g_{\mu \nu}$), we obtain the Einstein equation: 
\begin{align}\label{3deom}
  {\rm EOM}_{\mu \nu}: \  R_{\mu\nu}-\frac{R}{2}g_{\mu\nu}+\Lambda g_{\mu\nu}=0.
\end{align}
Next we compute the Neumann boundary condition (NBC) by varying (\ref{action-YBD}) with respect to the metric $h_{ij}$ as follows
\begin{eqnarray}
\label{NBC}
{\cal K}_{ij}^{\gamma}-({\cal K}^{\gamma}-T^{\gamma})h_{ij}^{\gamma}=0,
\end{eqnarray}
where ${\cal K}_{ij}^{\gamma}=\frac{1}{2} n^\theta \partial_\theta {g_{ij}^{\gamma}}|_{\theta=\theta_{1},\pi-\theta_{2}}$\footnote{Where $\theta$ is the location of the KR brane and $n^{\theta}$ is the unit normal to the KR brane. The unit normal is define through the following relation 
\begin{align}\label{unitvec}
    g_{MM}n^{M}n^{N}=1.
\end{align}
}; $\theta=\theta_{1},\pi-\theta_{2}$ denote the locations of the two Karch-Randall (KR) branes, see Fig. \ref{WHS}.
 Vacuum solution of the bulk Einstein's equation (\ref{3deom}) of the action (\ref{action-YBD}) is given by \cite{YB-AdS3}\footnote{This solution has been constructed from two-dimensional Yang-Baxter deformed JT gravity by uplifting two-dimensional solution to three dimensions, for more details, see \cite{YB-AdS3}.}:
\begin{eqnarray}
\label{metric-vacuum}
& & ds^2_\eta={\cal F}_\eta(z)\Biggl(\frac{-dt^2+dz^2}{z^2}\Biggr) +{\cal G}_{\theta \theta}^\eta d\theta^2,
\end{eqnarray}
where
\begin{eqnarray}
\label{F-G}
& &
{\cal F}_\eta(z)= \frac{1}{1-\frac{\eta^2 \alpha^2}{z^2}}; \ \ \ \  {\cal G}_{\theta \theta}^\eta = \Biggl[\frac{1}{2 \eta}\log\Biggl(\frac{1+ \frac{\eta \alpha}{z}}{1- \frac{\eta \alpha}{z}}\Biggr)+1 \Biggr]^2,
\end{eqnarray}
where $\eta$ is the Yang-Baxter deformation parameter. 
 The black hole solution is given by:
\begin{eqnarray}
\label{metric-BH}
ds^2_\eta={\cal F}_\eta(z)\Biggl(\frac{-f(z)dt^2+\frac{dz^2}{f(z)}}{z^2 }\Biggr) +{\cal G}_{\theta \theta}^\eta d\theta^2,
\end{eqnarray}
where $f(z)=1-\frac{z^2}{z_h^2}\hspace{1mm}$. Here, $z_h$ denotes the location of the black hole horizon. The black hole solution defined in (\ref{metric-BH}) has the following thermal entropy:
\begin{eqnarray}
\label{Thermal-Entropy}
& &  \hskip -0.6in
S_{\rm thermal}=\frac{A_{z=z_h}}{4 G_N^{(3)}}=\frac{\int_{\theta_1}^{\pi-\theta_2} d\theta \sqrt{{\cal G}_{\theta \theta}^\eta}|_{z=z_h}}{4 G_N^{(3)}}=\frac{\left(\pi-(\theta_2 +\theta_1)\right) {\sqrt{{\cal G}_{\theta \theta}^\eta}|_{z=z_h}}}{4 G_N^{(3)}}\nonumber\\
& & =-\frac{\left(\pi-(\theta_2 +\theta_1)\right) \alpha f'(z_h)}{8G_N^{(3)} }=\frac{\left(\pi-(\theta_2 +\theta_1)\right) \alpha }{4 z_h G_N^{(3)} }.
\end{eqnarray}
{\bf Wedge holography in the presence of homogeneous Yang-Baxter deformation}:
%\label{WHIHYB}
Wedge holography in three dimensional bulk exists provided following conditions should be satisfied \cite{Miao:2020oey,Akal:2020wfl}:
\begin{itemize}
\item Bulk metric should satisfy Einstein's equation with a negative cosmological constant in three dimensions (\ref{3deom}).

\item Bulk metric should satisfy Neumann boundary conditions (\ref{NBC}) on the Karch-Randall branes.

\item Metric ($h_{ij}$) on Karch-Randall branes should satisfy Einstein's equation with a negative cosmological constant in two dimensions:
\begin{align}\label{2deom}
    R_{ij}-\frac{R[h_{ij}]}{2}h_{ij}+\Lambda h_{ij}=0.
\end{align}

\end{itemize}
Wedge holography has been discussed in the context of compact angle in \cite{GB-3}. Let us briefly review the same here. The bulk metric has the following form:
\begin{eqnarray}
\label{Bulk-mu}
ds^2=\frac{1}{\sin^2\mu}\left(\frac{-dt^2+du^2+d\vec{x}^2}{u^2}+d\mu^2\right),
\end{eqnarray}
where $u>0$, $\vec{x}=(x^1,x^2,......,x^{d-2})$ [$d-2$ real transverse directions] and $\mu \in (0,\pi)$. In the bulk (\ref{Bulk-mu}), Karch-Randall branes are located at $\mu = \theta_1$ and $\mu = \pi-\theta_2$ [see Fig. {\bf 1} of \cite{GB-3}]. The geometry on the Karch-Randall branes is $AdS_d$. The setup constructed by us has the similar scenario. In our case, the compact direction is $\theta$, and the Karch-Randall branes have $AdS_2^\eta$ geometry. The difference between \cite{GB-3} and our setup is the absence of $d-2$ transverse direction in this work. For the black hole case, there will be black hole function in the metric (\ref{Bulk-mu}):
\begin{eqnarray}
\label{Bulk-mu-BH}
ds^2=\frac{1}{\sin^2\mu}\left(\frac{-h(u)dt^2+\frac{du^2}{h(u)}+d\vec{x}^2}{u^2}+d\mu^2\right),
\end{eqnarray}
where $h(u)=1-\frac{u^{d-1}}{u_h^{d-1}}$. \par
Wedge holography in the presence of homogeneous Yang-Baxter deformation can be constructed as follows. We have two Karch-Randall branes located at $\theta=\theta_{1}$, $\theta=\pi-\theta_{2}$ and these branes are joined at the one-dimensional defect $(P)$ as shown in Fig. \ref{WHS}. In this setup, the non-conformal boundary\footnote{$\eta$ deformation breaks the scale invariance therefore the boundary theory is not a CFT, see \cite{Roychowdhury:2017oqd}.} of the $3D$ bulk is located at the holographic screen placed at $z=z_B$\footnote{See \cite{HS-Chethan} where the holographic screen has been used in the literature.}. The homogeneous Yang-Baxter deformed wedge holography has the following three descriptions.

\begin{itemize}
\item {\bf Boundary description}: deformed SYK model living at the corner of wedge formed by two Karch-Randall branes with geometry homogeneous Yang-Baxter deformed AdS$_2$.

\item {\bf Intermediate description}: gravitating systems with homogeneous Yang-Baxter deformed $AdS_2$ geometries are connected to each other via transparent boundary conditions at the defect.

\item {\bf Bulk description}: the holographic dual of deformed SYK model is three dimensional gravity living in the wedge formed by Karch-Randall branes.
\end{itemize}

\par
In our setup, the dictionary of wedge holography in the presence of homogeneous Yang-Baxter deformation can be schematically expressed as:\\ \\
\fbox{\begin{minipage}{38em}{\it 
Classical gravity in three-dimensional homogeneous Yang-Baxter deformed $AdS_3$ bulk\\ $\equiv$ (Quantum) gravity on two Karch-Randall branes with metric homogeneous Yang-Baxter deformed $AdS_2$\\ $\equiv$ deformed SYK model living at the one-dimensional defect.}
\end{minipage}}\\  \\
The relationship between the first and second lines is because of the braneworld holography \cite{KR1,KR2}. The third line is related to the second line due to JT/SYK duality \cite{Sachdev:1992fk}. Therefore, {\it classical gravity in homogeneous Yang-Baxter deformed $AdS_3$ can be dual to deformed SYK model at the one-dimensional defect} based on wedge holography \cite{WH-i,WH-ii,WH-iii}.

{\bf Tensions of Karch-Randall branes without DGP term}: In this case, bulk action is (\ref{action-YBD}). For general metric, NBC (\ref{NBC}) is satisfied provided, $T^{\gamma}={\cal K}^{\gamma}-\frac{{\cal K}_{ij}^{\gamma}}{h_{ij}^{\gamma}}$. For vacuum solution (\ref{metric-vacuum}) and black hole solutin (\ref{metric-BH}), $K_{ij}^{\gamma}=0$, and hence ${\cal K}^{\gamma}=0$, therefore (\ref{NBC}) implying 
\begin{eqnarray}
\label{NBC-i}
T^{\gamma}h_{ij}^{\gamma}=0.
\end{eqnarray}
\par
Notice that, $h_{ij}^{\gamma} \neq 0$, and hence $T_\gamma=0$, i.e., Karch-Randall branes are tensionless without DGP term. \\

{\bf Tensions of Karch-Randall branes in the presence of DGP term}: In the presence of the DGP term on the Karch-Randall branes, gravitational action (\ref{action-YBD}) is modified as \cite{Massless-Gravity}:
{\footnotesize
\begin{eqnarray}
\label{bulk-action-DGP}
& &
\hskip -0.2in S=\frac{1}{16 \pi G_N^{(3)}}\Biggl[ \int_M d^{3}x \sqrt{-g^{(3)}} \left(R[g] -2 \Lambda\right)+2\int_{\partial M} d^2 x\sqrt{-h}K+2 \int_{Q_\gamma} d^2x \sqrt{-h_\gamma}\left({\cal K}_\gamma-T_\gamma+\lambda_\gamma R_{h_\gamma}\right)\Biggr] , 
\end{eqnarray}
}
where all the terms in (\ref{bulk-action-DGP}) are same as defined in (\ref{action-YBD}) except $R_{h_\gamma}$, which are intrinsic curvature scalars on two Karch-Randall branes. In this case, bulk metric satisfies the following Neumann boundary condition at $\theta=\theta_{1},\pi-\theta_{2}$:
\begin{eqnarray}
\label{NBC-DGP}
{\cal K}_{\gamma,ij}-({\cal K}_\gamma-T_\gamma+\lambda_\gamma R_{h_\gamma} )h_{\gamma,ij}+2 \lambda_\gamma R_{h_\gamma, ij} =0.
\end{eqnarray}
\par
The general bulk metric satisfies Neumann boundary condition (\ref{NBC-DGP}) provided tensions of the Karch-Randall branes in the presence of DGP term should be given as follows: 
\begin{eqnarray}
\label{tensions-branes-DGP}
T_\gamma={\cal K}_{\gamma}-\frac{{\cal K}_{\gamma,ij}}{h_{\gamma,ij}}+\lambda_\gamma R_{h_\gamma}-\frac{2 \lambda_\gamma R_{h_\gamma, ij}}{h_{\gamma,ij}}.
\end{eqnarray}
For the bulk metric (\ref{metric-vacuum}) and on $\theta=\theta_{1},\pi-\theta_{2}$ slices, one can show that
\begin{eqnarray}
\label{T1-T2}
& &
T_1=2 {\lambda_1}+\frac{8 \kappa ^2 {\lambda_1}}{z^2}-\frac{2 \kappa ^4 {\lambda_1}}{z^4}+O\left(\kappa ^6\right);\ 
T_2=2 {\lambda_2}+\frac{8 \kappa ^2 {\lambda_2}}{z^2}-\frac{2 \kappa ^4 {\lambda_2}}{z^4}+O\left(\kappa ^6\right),
\end{eqnarray}
where $\kappa=\eta \alpha$.

\section{Information paradox in homogeneous Yang-Baxter deformed wedge holography}
\label{IP}
In this section, we use the wedge holography constructed in section \ref{WH-YBAdS} to discuss the Page curve of black holes in the homogeneous Yang-Baxter deformed wedge holography in \ref{IP-WDGP}.
 and in \ref{IP-WDGP-i} we discuss the DGP term and swampland criteria in homogeneous Yang-Baxter deformed wedge holography.
\subsection{Page curve due to the presence of homogeneous Yang-Baxter deformation}
\label{IP-WDGP}
In this subsection, we explore the Page curve of the eternal black hole in the absence of the DGP term on the Karch-Randall branes. For this purpose, we consider the two extremal surfaces: Hartman-Maldacena, and island surfaces, and calculate their respective entanglement entropies in \ref{HM-WDGP} and \ref{IS-WDGP}.
\subsubsection{Hartman-Maldacena surface's entanglement entropy}
\label{HM-WDGP}
 Let us write the bulk metric (\ref{metric-BH}) in terms of the infalling Eddington-Finkelstein coordinate, $d v= dt -\frac{dz}{f(z)}$ as below:
\begin{eqnarray}
\label{metric-bulk-BH-EFC}
ds_{(2+1)}^2=g_{\mu \nu} dx^\mu dx^\nu={\cal G}_{\theta \theta}^\eta d\theta^2+{{\cal F}_\eta(z)}\Biggl(\frac{-f(z) dv^2-2 dv dz }{z^2}\Biggr).
\end{eqnarray}
Hartman-Maldacena surface is parametrized by $\theta \equiv \theta(z)$ and $v \equiv v(z)$ which has the induced metric as given below, and the same can be obtained from (\ref{metric-bulk-BH-EFC}):
\begin{eqnarray}
\label{induced-metric-HM}
& & ds^2= \Biggl({\cal G}_{\theta \theta}^\eta{\theta'(z)^2-\frac{{{\cal F}_\eta(z)}v'(z)}{z^2} \left(2+f(z)v'(z)\right)}\Biggr) dz^2,
\end{eqnarray}
where $\theta'(z)=\frac{d\theta}{dz}$ and $v'(z)=\frac{dv}{dz}$. Using (\ref{induced-metric-HM}), the area of the Hartman-Maldacena surface can be computed as:
\begin{eqnarray}
\label{AHM-WDGP}
& & A_{\rm HM}= \int_{z_1}^{z_{\rm max}} dz \Biggl(  \sqrt{ {\cal G}_{\theta \theta}^\eta{\theta'(z)^2-\frac{{{\cal F}_\eta(z)}v'(z)}{z^2} \left(2+f(z)v'(z)\right)}}\Biggr),
\end{eqnarray}
where $z_1$ and $z_{\rm max}$ are the point on gravitating bath and turning point of the Hartman-Maldacena surface. From (\ref{AHM-WDGP}), we obtain the equation of motion for $\theta(z)$ as
\begin{eqnarray}
\label{EOM-theta(z)}
\frac{{\cal G}_{\theta \theta}^\eta \theta '(z)}{\sqrt{{\cal G}_{\theta \theta}^\eta \theta '(z)^2-\frac{{\cal F}_\eta(z) v'(z) \left(f(z)
   v'(z)+2\right)}{z^2}}}=C_1,
\end{eqnarray}
where $C_1$ is a constant. Solving the above equation for $\theta'(z)$, we obtained:
\begin{eqnarray}
\label{soln-theta(z)}
\theta'(z)=\frac{C_1 \sqrt{{\cal F}_\eta(z)} \sqrt{v'(z)} \sqrt{f(z) v'(z)+2}}{\sqrt{{\cal G}_{\theta \theta}^\eta} z
   \sqrt{C_1^2-{\cal G}_{\theta \theta}^\eta}},
\end{eqnarray}
 Substituting $\theta'(z)$ from (\ref{soln-theta(z)}) into (\ref{AHM-WDGP}), we obtained:
\begin{eqnarray}
\label{AHM-WDGP-i}
& & \hskip -0.3in A_{\rm HM}= \int_{z_1}^{z_{\rm max}} dz {\cal L}_{\rm HM}= \int_{z_1}^{z_{\rm max}} dz \Biggl(  \sqrt{\frac{{\cal F}_\eta(z) v'(z) \left(f(z) v'(z) \left(C_1^2 {\cal F}_\eta(z)+z^2
   \left({\cal G}_{\theta \theta}^\eta-C_1^2\right)\right)+2 {\cal G}_{\theta \theta}^\eta z^2\right)}{z^4
   \left(C_1^2-{\cal G}_{\theta \theta}^\eta\right)}}\Biggr). \nonumber\\
   & &
\end{eqnarray}
The equation of motion for the embedding $v(z)$ from (\ref{AHM-WDGP-i}) is calculated as
{
\begin{eqnarray}
\label{abc}
& & 
\frac{d}{dz}\left(\frac{\partial {\cal L}_{\rm HM}}{\partial v'(z)}\right)=0 \ \ \implies \frac{\partial {\cal L}_{\rm HM}}{\partial v'(z)}=E.
\end{eqnarray}
}
Using (\ref{AHM-WDGP-i}) and (\ref{abc}), we obtained:
\begin{eqnarray}
\label{E}
E=\frac{{\cal F}_\eta(z) \left(f(z) v'(z) \left(C_1^2 {\cal F}_\eta(z)+z^2
   \left({\cal G}_{\theta \theta}^\eta-C_1^2\right)\right)+{\cal G}_{\theta \theta}^\eta z^2\right)}{z^4
   \left(C_1^2-{\cal G}_{\theta \theta}^\eta\right) \sqrt{\frac{{\cal F}_\eta(z) v'(z) \left(f(z) v'(z) \left(C_1^2 {\cal F}_\eta(z)+z^2
   \left({\cal G}_{\theta \theta}^\eta-C_1^2\right)\right)+2 {\cal G}_{\theta \theta}^\eta z^2\right)}{z^4
   \left(C_1^2-{\cal G}_{\theta \theta}^\eta\right)}}}.
\end{eqnarray}
Using the condition $v'(z_{\rm max}) \rightarrow - \infty$ \cite{NGB}, equations (\ref{F-G}) and (\ref{E}), we obtained: 
%$E$ at $z_{\rm max}$ as follows:
%\begin{eqnarray}
%\label{Ezmax}
%E=\sqrt{\frac{\left(1-z_{\rm max}^2\right) \left(z_{\rm max}^2 \left(\frac{\left(2 \eta +\log \left(-\frac{\kappa
 %  +z_{\rm max}}{\kappa -z_{\rm max}}\right)\right)^2}{4 \eta ^2}-C_1^2\right)+\frac{C_1^2}{1-\frac{\kappa
%   ^2}{z_{\rm max}^2}}\right)}{z_{\rm max}^4 \left(1-\frac{\kappa ^2}{z_{\rm max}^2}\right) \left(C_1^2-\frac{\left(2 \eta
 %  +\log \left(-\frac{\kappa +z_{\rm max}}{\kappa -z_{\rm max}}\right)\right)^2}{4 \eta ^2}\right)}}.
%\end{eqnarray}
\begin{eqnarray}
\label{Ezmax-kappa}
E=\sqrt{\frac{\left(z_{\rm max}^2-1\right) \left(C_1^2
   z_{\rm max}^2-C_1^2-z_{\rm max}^2\right)}{\left(C_1^2-1\right) z_{\rm max}^4}}-\frac{C_1^2 \kappa  \sqrt{\frac{\left(z_{\rm max}^2-1\right) \left(C_1^2
   z_{\rm max}^2-C_1^2-z_{\rm max}^2\right)}{\left(C_1^2-1\right) z_{\rm max}^4}}}{\left(C_1^2-1\right) \eta 
   z_{\rm max} \left(C_1^2 z_{\rm max}^2-C_1^2-z_{\rm max}^2\right)}.
\end{eqnarray}
where we retain the terms upto ${\cal O}(\kappa)$. At the turning point, $\frac{d E}{dz_{\rm max}}=0$ which implies $z_{\rm max}=\frac{\sqrt{2} C_1}{\sqrt{2 C_1^2-1}}-\left(\sqrt{\frac{-3 C_1^2+\sqrt{4-3 C_1^2}+2}{3-3 C_1^2}}\right) \kappa$. To make sure that the turning point of the Hartman-Maldacena surface is outside the horizon and positive, we need to consider $C_1 > \frac{1}{\sqrt{2}}$ (e.g., for $C_1=\frac{1}{\sqrt{1.6}}$, $z_{\rm max} \approx 2.2 - 1.2 \kappa$). Time on the gravitating bath can be obtained after simplifying $d v= dt -\frac{dz}{f(z)}$ and the same is written below:
\begin{eqnarray}
\label{tz1}
t_1=t(z_1)=-\int_{z_1}^{z_{\rm max}} \left(v'(z)+\frac{1}{f(z)}\right)dz.
\end{eqnarray}
Now we explore the late-time behavior ($t \rightarrow \infty$) of the area of the Hartman-Maldacena surface as follows:
\begin{eqnarray}
& & {\rm lim}_{t \rightarrow \infty} \frac{dA_{\rm HM}}{dt}={\rm lim}_{t \rightarrow \infty}\Biggl( \frac{\frac{dA_{\rm HM}}{dz_{\rm max} }}{\frac{dt}{dz_{\rm max} }}\Biggr) = \frac{{\cal L}_{\rm HM}(z_{\rm max},v'(z_{\rm max}))+\int_{z_1}^{z_{\rm max}} \frac{\partial {\cal L}_{\rm HM}(z_{\rm max},v'(z_{\rm max}))}{\partial z_{\rm max} }dz}{-v'(z_{\rm max})-\frac{1}{f(z_{\rm max})}-\int_{z_1}^{z_{\rm max}}\frac{\partial v'(z)}{\partial z_{\rm max} }}.
\end{eqnarray}
Since,
\begin{eqnarray}
& & \hskip -0.6in
{\rm lim}_{t \rightarrow \infty}\frac{\partial v'(z)}{\partial z_{\rm max} }={\rm lim}_{t \rightarrow \infty}\frac{\partial v'(z)}{\partial E}\frac{\partial E}{\partial z_{\rm max} }=0; \ \ {\rm lim}_{t \rightarrow \infty}\frac{\partial L(z,v'(z))}{\partial z_{\rm max} } = \frac{\partial {\cal L}_{\rm HM}(z,v'(z))}{\partial v'(z) } \frac{\partial v'(z)}{\partial z_{\rm max} }=0.
\end{eqnarray}
Therefore,
\begin{eqnarray}
{\rm lim}_{t \rightarrow \infty} \frac{dA_{\rm HM}}{dt} = \frac{{\cal L}_{\rm HM}(z_{\rm max},v'(z_{\rm max}))}{-v'(z_{\rm max})-\frac{1}{f(z_{\rm max})}}   = Constant.
\end{eqnarray}

Hence, the Hartman-Maldancena surface has the following entanglement entropy in Yang-Baxter deformed model:
\begin{eqnarray}
\label{SHM}
S_{\rm HM}=\frac{A_{\rm HM}}{4 G_N^{(3)}} \propto t.
\end{eqnarray}

\subsubsection{Non-trivial island surface in the presence of homogeneous Yang-Baxter deformation} 
\label{IS-WDGP}
Island surface is parametrized by $t =constant$, $z=z(\theta)$, and hence induced metric of the same can be obtained from (\ref{metric-BH}) as written below:
\begin{eqnarray}
\label{IM-Island}
ds^2_{\rm Island}=\Biggl(\frac{{\cal F}_\eta(z) z'(\theta)^2}{z^2 f(z)}+{\cal G}_{\theta \theta}^\eta  \Biggr)d\theta^2.
\end{eqnarray}
\par
The area of the island surface using the induced metric (\ref{IM-Island}) is obtained as follows:
\begin{eqnarray}
\label{AIS-YB-AdS}
& &
{\cal A}_{\rm Island} =\int_{\theta_1}^{\pi-\theta_2} d\theta \Biggl(\sqrt{\frac{{\cal F}_\eta(z) z'(\theta)^2}{z^2 f(z)}+{\cal G}_{\theta \theta}^\eta }\Biggr).
\end{eqnarray}
Using (\ref{F-G}), and $f(z)=1-\frac{z^2}{z_h^2}$(we used $z_h=1$ for the simplicity of the calculations) and (\ref{AIS-YB-AdS}), we obtained:
\begin{eqnarray}
\label{AIS-YB-AdS-simp}
& & 
{\cal A}_{\rm Island} =\int_{\theta_1}^{\pi-\theta_2} d\theta \Biggl(\sqrt{\frac{1}{4 \eta ^2}\left(2 \eta +\log \left(-\frac{\kappa +z(\theta )}{\kappa -z(\theta )}\right)\right)^2+\frac{z'(\theta
   )^2}{\left(z(\theta )^2-1\right) \left(\kappa ^2-z(\theta )^2\right)}}\Biggr).
\end{eqnarray}
\par
The Euler-Lagrange equation of motion(EOM) for the island surface's embedding $z(\theta)$ upto ${\cal O}(\kappa^2)$ can be schematically expressed as
\begin{eqnarray}
\label{EOM-kappa-expansion}
{\rm EOM}={\rm EOM}^{\kappa^0} - \left({\rm EOM}^{\kappa^1}\right) \kappa - \left({\rm EOM}^{\kappa^2}\right) \kappa^2,
\end{eqnarray}
where we defined ${\rm EOM}^{\kappa^{0,1,2}}$ as below:
\begin{itemize}
\item {\bf ${\rm EOM}^{\kappa^0}$}:
\begin{eqnarray}
\label{EOM-i}
& & 8 \Biggl(\eta ^2 z(\theta )^{10} z''(\theta )-2 \eta ^2 z(\theta )^8 z''(\theta )+\eta ^2 z(\theta )^6 z''(\theta )-2 \eta ^2
   z(\theta )^9 z'(\theta )^2+3 \eta ^2 z(\theta )^7 z'(\theta )^2 \nonumber\\
   & & -\eta ^2 z(\theta )^5 z'(\theta )^2\Biggr)=0.
\end{eqnarray}

\item {\bf ${\rm EOM}^{\kappa^1}$}:
\begin{eqnarray}
\label{EOM-ii}
& & 8 \Biggl(-2 \eta  z(\theta )^9 z''(\theta )+4 \eta  z(\theta )^7 z''(\theta )-2 \eta  z(\theta )^5 z''(\theta )+2 \eta  z(\theta
   )^8 z'(\theta )^2-2 \eta  z(\theta )^6 z'(\theta )^2 \nonumber\\
   & &+\eta  z(\theta )^{12} -3 \eta  z(\theta )^{10}+3 \eta  z(\theta )^8-\eta 
   z(\theta )^6\Biggr)=0.
\end{eqnarray}

\item {\bf ${\rm EOM}^{\kappa^2}$}:
\begin{eqnarray}
\label{EOM-iii}
& & 8 \Biggl(3 \eta ^2 z(\theta )^8 z''(\theta )-6 \eta ^2 z(\theta )^6 z''(\theta )+3 \eta ^2 z(\theta )^4 z''(\theta )-5 \eta ^2
   z(\theta )^7 z'(\theta )^2+7 \eta ^2 z(\theta )^5 z'(\theta )^2
   \nonumber\\
   & &
   -2 \eta ^2 z(\theta )^3 z'(\theta )^2-z(\theta )^8 z''(\theta
   )+2 z(\theta )^6 z''(\theta )-z(\theta )^4 z''(\theta )+z(\theta )^5 z'(\theta )^2-z(\theta )^3 z'(\theta )^2\nonumber\\
   & &+3 z(\theta
   )^{11}
   -9 z(\theta )^9+9 z(\theta )^7-3 z(\theta )^5\Biggr)=0.
\end{eqnarray}

\end{itemize}
EOMs (\ref{EOM-i}), (\ref{EOM-ii}), and (\ref{EOM-iii}) have a common physical solution which is given below. 
\begin{equation}
\label{common-soln}
z(\theta)=1
\end{equation}
\par
As discussed earlier, $z_h=1$ is the black hole horizon. Therefore solution to the island surface's embedding $z(\theta)$ up to ${\cal O}(\kappa^2)$ is given by:
\begin{eqnarray}
\label{soln-IS-kappa^2}
z(\theta)=1-\kappa-\kappa^2 = z^{\rm YB \ deformed}
\end{eqnarray}
\begin{figure}
\begin{center}
\includegraphics[width=0.5\textwidth]{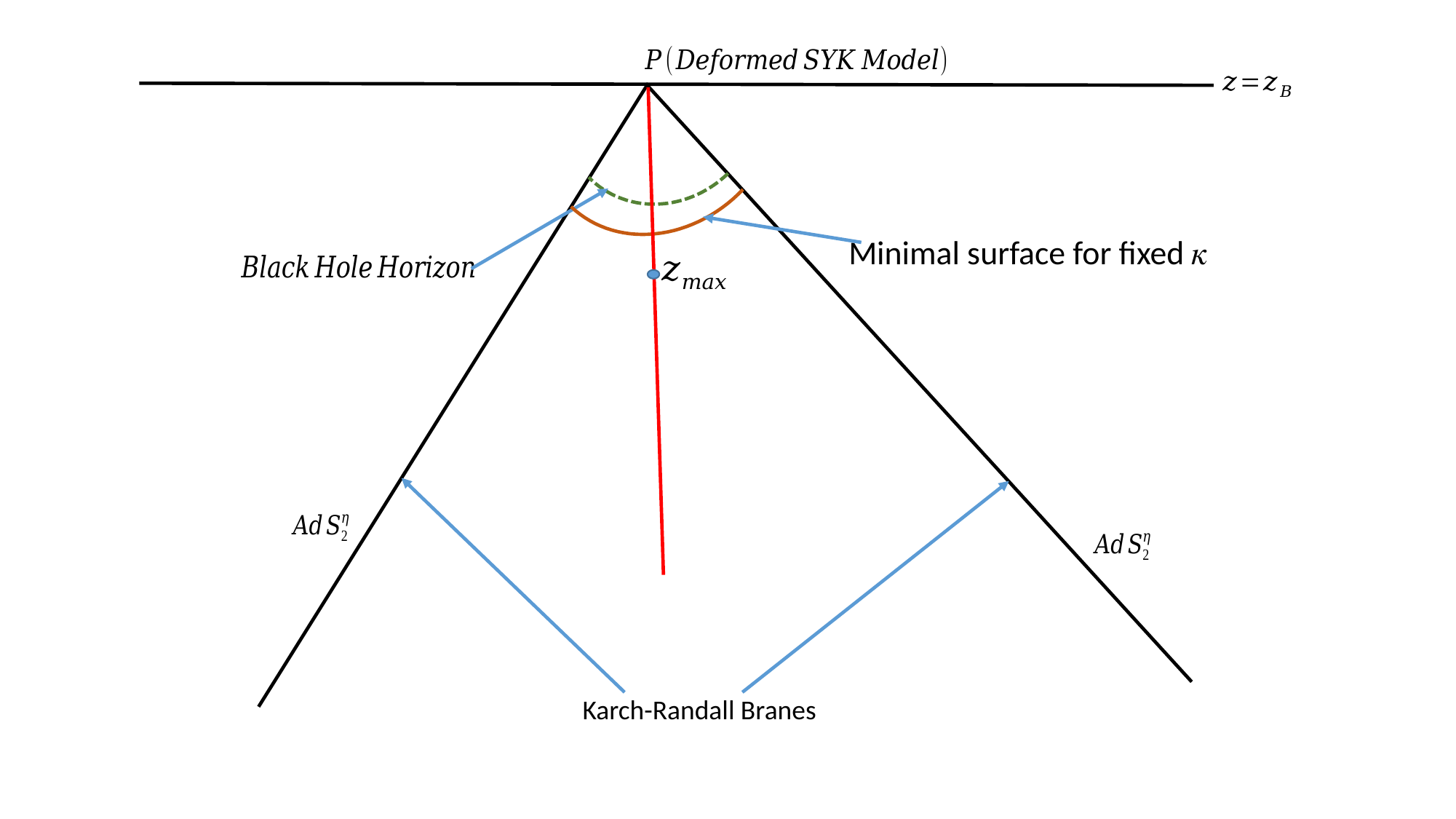}
\end{center}
\caption{Description of homogeneous Yang-Baxter deformed wedge holography. The presence of homogeneous Yang-Baxter deformation makes it possible to have a non-trivial island surface (solid brown curve) inside the horizon. The green dotted line in the figure corresponds to the black hole horizon. {In this figure, we have shown just one part of the wedge holography; the Hartman-Maldacena surface will join the defect located on the other side of wedge holography}.}
\label{WHS-IP}
\end{figure}
We can see very easily that (\ref{soln-IS-kappa^2}) satisfies the NBC on the branes, i.e., $z'(\theta)|_{\theta=\theta_{1},\pi-\theta_{2}}=0$. For $0<\kappa<1$, equation (\ref{soln-IS-kappa^2}) implying $z(\theta)<1$, i.e., $z^{\rm YB \ deformed}<z_h$. Since $z(\theta)$ is constant, therefore substituting $z(\theta)$ from (\ref{soln-IS-kappa^2}) into (\ref{AIS-YB-AdS}), we get the minimal area of the island surface as:
\begin{eqnarray}
\label{AIS-YB-AdS-LT}
& & 
{\cal A}_{\rm Island} =\int_{\theta_1}^{\pi-\theta_2} d\theta \sqrt{{\cal G}_{\theta \theta}^\eta|_{z=z^{\rm YB \ deformed}} }.
\end{eqnarray}
\par
According to Ryu-Takayanagi's prescription the entanglement entropy of the island surfaces is obtained as \cite{RT}:
\begin{eqnarray}
\label{SIS}
& &
S_{\rm Island}=\frac{2{\cal A}_{\rm Island}}{4 G_N^{(3)}}=  \frac{2 \int_{\theta_1}^{\pi-\theta_2} d\theta \sqrt{{\cal G}_{\theta \theta}^\eta|_{z^{\rm YB \ deformed}} } } {4 G_N^{(3)}} \nonumber\\
& &
= -\frac{2 \left(\pi-(\theta_2 +\theta_1)\right) \alpha f'(z^{\rm YB \ deformed}) } {8 G_N^{(3)}} = 2\left(\frac{ \left(\pi-(\theta_2 +\theta_1)\right) \alpha z^{\rm YB \ deformed} } {4 G_N^{(3)}}\right)
<2 S_{\rm thermal}.\nonumber\\
\end{eqnarray}
The factor ``2'' in the above equation is due to the second island surface contribution from the thermofield double partner. Therefore, we get a {\it non-trivial island surface} inside the horizon due to the presence of homogeneous Yang-Baxter deformation whose entanglement entropy is lower than twice of thermal entropy of the black hole.

\subsubsection{Page curve}
\begin{figure}
\begin{center}
\includegraphics[width=0.5\textwidth]{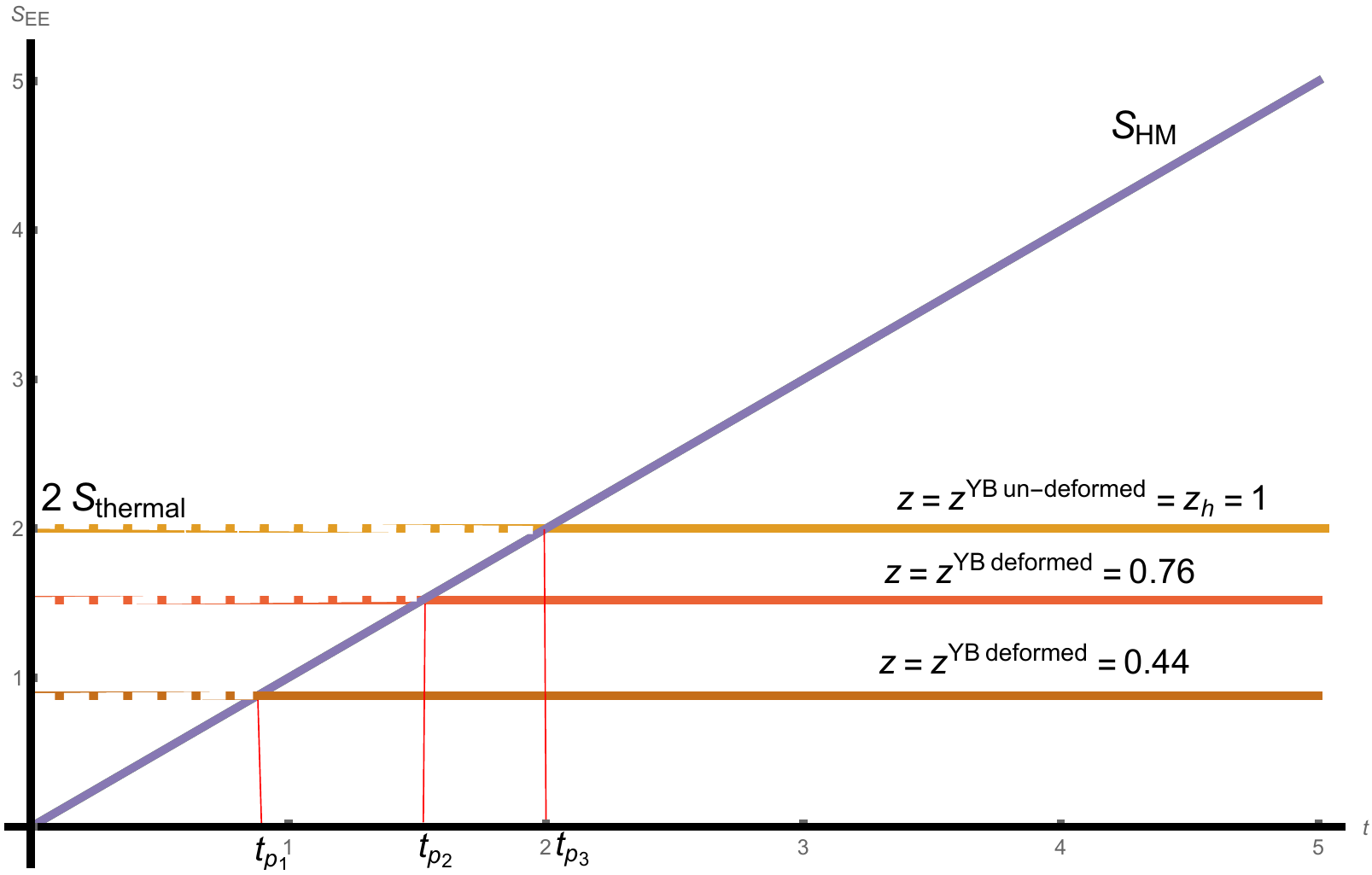}
\end{center}
\caption{Page curves for various values of $\kappa$.}
\label{Page-Curves}
\end{figure}

From subsections \ref{HM-WDGP} and \ref{IS-WDGP}, we see that the Hartman-Maldacena surface has the linear time dependence (\ref{SHM}) and island surface's entanglement entropy is less than  the twice of thermal entropy of the black hole (\ref{SIS}), hence the Page curve without DGP term is obtained as shown in Fig. \ref{Page-Curves}.
Fig. \ref{Page-Curves} has been drawn by dropping the overall factor $\left(\frac{\left(\pi-(\theta_2 +\theta_1)\right) \alpha  } {4 G_N^{(3)}}\right)$. This factor is also common in the thermal entropy of the black hole (\ref{Thermal-Entropy}). If we consider the numerical value of the aforementioned factor, then this will just scale the Page curves, but the qualitative results will remain the same. To draw the figure, we have taken $\kappa=0, 0.2, 0.4$ implying $z^{\rm YB \ un-deformed}=z_h=1$ and $z^{\rm YB \ deformed}=0.76, 0.44$ respectively. The entanglement entropies for $z^{\rm YB \ deformed}=0.76, 0.44$ are $S_{\rm Island} =1.52, 0.88$ which are less than $2 S_{\rm thermal} = 2$. \par

Let us understand the physical meaning of these results. Since $z^{\rm YB \ deformed}<z_h$ therefore, extremal surfaces in the presence of homogeneous Yang-Baxter deformation are behind the black hole horizon. See \cite{Gopal+Nitin,I-3} where the island surface was found inside the horizon in non-holographic models. We found this result for the holographic model in this work. Using the entanglement entropies of the island surface and Hartman-Maldacena surface, we obtained the Page curve in wedge holography in the presence of homogeneous Yang-Baxter deformation.

Based on \cite{scrambling-time-1,scrambling-time-2,scrambling-time-EW}, we can say that as soon as the island comes into the picture, one starts recovering the information from the black hole. In Fig. \ref{Page-Curves}, we see that we get different Page curves for different values of $\kappa$, and hence the homogeneous Yang-Baxter deformation is shifting the Page curves. More explicitly, when $\kappa$ increase, then Page curves shift towards earlier time. Therefore, variation of $\kappa$ affects the emergence of islands and information recovery of the Hawking radiation\footnote{See \cite{RNBH-HD,Ankit} where the higher derivative terms affect the Page curves.}.

\subsection{DGP term and swampland criteria in homogeneous Yang-Baxter deformed wedge holography}
\label{IP-WDGP-i}
In this subsection, we will discuss the effect of including the DGP term on the Karch-Randall branes. As discussed in \cite{Massless-Gravity}, the Hartman-Maldacena surface entanglement entropy remains same. In the presence of the DGP term, the island surface's entanglement entropy 
receives an extra contribution from the boundary of the island surface \cite{Massless-Gravity}. In $AdS_{d+1}/CFT_d$ correspondence, the same can be expressed as follows:
\begin{eqnarray}
\label{SEE-WDGP}
S_{EE} \sim {\rm min} \Biggl[{\rm ext}\biggl( \int_{\Gamma} d^{d-1}x \sqrt{\gamma}+ \int_{\partial \Gamma} d^{d-2}x \sqrt{\sigma} \lambda_a\Biggr)\Biggr],
\end{eqnarray}
where $\Gamma$ being the RT surface with induce metric $\gamma$ and boundary of $\Gamma$ ($\partial \Gamma$) has the induced metric $\sigma$. 
\par
Recently it has been pointed out that including DGP term on the Karch-Randall brane is non-physical \cite{DGP-Swampland,DGP-Swampland-i}. DGP term  can leads to negative entanglement entropy because of negative effective coupling constant of one brane. To have a positive entanglement entropy in the homogeneous Yang-Baxter deformed wedge holography we need to consider $(\theta_1+\theta_2) < \pi$. Another swampland criteria mentioned by authors in \cite{DGP-Swampland} is that for any extremal surface if $S_{\rm ext} < S_{\rm thermal}$ then those theories will belong to the swampland. 
\par
We obtain the non-trivial island surface in homogeneous Yang-Baxter deformed wedge holography due to the homogeneous Yang-Baxter deformation. We can have positive entanglement entropy for the island surface provided $(\theta_1+\theta_2) < \pi$ and hence we can avoid one of the swampland criteria given in \cite{DGP-Swampland}. But in our setup $S_{\rm island} < S_{\rm thermal}$ therefore the second criteria of swampland is unavoidable. Based on the results of \cite{DGP-Swampland} we can say that homogeneous Yang-Baxter deformed wedge holography is also the part of swampland.

\section{Holographic complexity in homogeneous Yang-Baxter deformed background}
\label{HC}
In this section, we compute the holographic complexity of homogeneous Yang-Baxter deformed $AdS_2$ using the complexity equals volume proposal \cite{CV}\footnote{Computation in this section is based on usual holography. We are not using wedge holography in this section.}. This proposal states that the complexity of holographic dual theory is given by the volume of co-dimension one surface in the bulk. For our case, we have ``$AdS_2^\eta$/deformed SYK'' duality, and hence we can write the expression for the holographic complexity as:
\begin{eqnarray}
& & \label{CV}
{\cal C}=\frac{{\cal V}}{G_N^{(2)} l},
\end{eqnarray}
where ${\cal V}$, $G_N^{(2)}$, and $l$ are the volume of a one-dimensional surface in two-dimensional bulk, Newton constant in two dimensions, and length scale associated with Yang-Baxter deformed $AdS_2$ background respectively.\par
Let us write the two dimensional metric for homogeneous Yang-Baxter deformed background as:
\begin{eqnarray}
\label{IB-YBAdS2-i}
ds^2=\frac{{\cal F}_\eta(z)}{z^2}\Biggl(-f(z)dt^2+\frac{dz^2}{f(z)}\Biggr),
\end{eqnarray}
where 
\begin{eqnarray}
\label{2d-f-F}
& & {\cal F}_\eta(z)=\frac{1}{1-\frac{\kappa^2}{z^2}}; \ \ \ f(z)=1-\frac{z^2}{z_h^2}.
\end{eqnarray}
In order to compute holographic complexity, we parametrize the volume slice by ``$t(z)$''. The induced metric associated with co-dimension one surface in homogeneous Yang-Baxter deformed $AdS_2$ is obatined using (\ref{IB-YBAdS2-i}) and it is written below:
\begin{eqnarray}
\label{IB-YBAdS2}
ds^2|_\gamma=\frac{{\cal F}_\eta(z)}{z^2}\Biggl(-f(z)t'(z)^2+\frac{1}{f(z)}\Biggr)dz^2.
\end{eqnarray}
Using (\ref{IB-YBAdS2}), the volume of co-dimension one surface is obtained as:
\begin{equation}
\label{V1}
{\cal V}=\int dz {\cal L}\left(z,t'(z)\right)=\int dz \left(\frac{\sqrt{{\cal F}_\eta(z)}\sqrt{\frac{1}{f(z)}-f(z) t'(z)^2}}{z }\right).
\end{equation}
Since there is no explicit $t(z)$ dependence in Lagrangian of (\ref{V1}) therefore energy will be conserved (say $E$), and is given as below:
\begin{eqnarray}
\label{CM-tz}
E=-\frac{\partial {\cal L}\left(z,t'(z)\right)}{\partial t'(z)}=\frac{f(z) \sqrt{{\cal F}_\eta(z)} t'(z)}{z \sqrt{\frac{1}{f(z)}-f(z) t'(z)^2}}.
\end{eqnarray}
On solving the above equation for $ t'(z)$, we obtained:
\begin{eqnarray}
\label{t'(z)}
t'(z)=\frac{E z}{f(z) \sqrt{E^2 z^2+f(z) {\cal F}_\eta(z)}}.
\end{eqnarray}
We can invert the above equation to get:
\begin{eqnarray}
\label{z'(t)}
z'(t)\equiv \frac{dz}{dt}=\frac{f(z) \sqrt{E^2 z^2+f(z) {\cal F}_\eta(z)}}{E z}.
\end{eqnarray}
\par
At the turning point $dz/dt|_{z=z_t}=0$, and substituting $f(z)$, ${\cal F}_\eta(z)$ from (\ref{2d-f-F}) into (\ref{z'(t)}), turning point from (\ref{z'(t)}) is obtained as:
\begin{eqnarray}
\label{zt}
z_t=\frac{{z_h} \sqrt{1-E^2 \kappa ^2}}{\sqrt{1-E^2 {z_h}^2}}
\end{eqnarray}
Now substituting $t'(z)$ from (\ref{t'(z)}) into the action (\ref{V1}) and utilizing $f(z)$, ${\cal F}_\eta(z)$ from (\ref{2d-f-F}), we obtained the on-shell volume as:
\begin{eqnarray}
\label{V2}
& & {\cal V}=\int_{z_B}^{z_t} {\cal L}=\int_{z_B}^{z_t}  \frac{ dz}{z}\left(\sqrt{\frac{z^2}{z^2-\kappa ^2}} \sqrt{\frac{{z_h}^2}{{z_h}^2 \left(1-E^2 \kappa ^2\right)+z^2 \left(E^2 {z_h}^2-1\right)}}\right).
\end{eqnarray}
 Utilizing (\ref{2d-f-F}) and (\ref{t'(z)}), we obtained the boundary time in the form of following integral:
\begin{eqnarray}
\label{tb}
& & 
t(z_B)=t_b=-\int_{z_B}^{z_t}dz\left(\frac{E z}{\left(1-\frac{z^2}{{z_h}^2}\right) \sqrt{E^2 z^2+\frac{1-\frac{z^2}{{z_h}^2}}{1-\frac{\kappa ^2}{z^2}}}} \right).
\end{eqnarray}
At late times, $E$ reaches a critical value, ``$E_{\rm crit}$'' which can be obtained by extremization of $E$ with respect to $z_t$ using (\ref{zt}) as follows:
\begin{eqnarray}
& & \frac{d E}{d z_t}=-\frac{\sqrt{1-E^2 \kappa ^2} \left(1-E^2
   {z_h}^2\right)^{3/2}}{E {z_h} \left(\kappa ^2-{z_h}^2\right)}.
\end{eqnarray}
Hence, extrimation of $E$ $\left(\frac{d E}{d z_t}|_{E=E_{\rm crit}}=0\right)$ leads to:
\begin{eqnarray}
\label{Ccrit}
E_{\rm crit}=\frac{1}{z_h}.
\end{eqnarray} 
Substituting $E_{\rm crit}$ from (\ref{Ccrit}) into (\ref{V2}) and using the definition of complexity as given in (\ref{CV}), we obtained the complexity growth at late times after differentiating obtained complexity with respect to boundary time ($t_b$) as given below:
\begin{eqnarray}
\label{dC-dtb}
& & 
{\frac{d {\cal C}}{d t_b}}\Biggl|_{t_b \rightarrow \infty}= \frac{\phi^{\rm YB} }{G_N^{(2)}lz_h}
\end{eqnarray}
where dilaton, $\phi^{\rm YB}={\sqrt{{\cal G}_{\theta \theta}^\eta}}|_{z=z_h}$, captures information about the transverse space. This can be understood from appendix \ref{3D-to-2D}. Utilizing the results obtained in appendix \ref{Thermodynamics-YBJT} for the Hawking temperature (\ref{Hawking-temp-YBJT}), we see that complexity growth at late times (\ref{dC-dtb}) simplified as:
\begin{eqnarray}
\label{dC-dtb-i}
& & 
{\frac{d {\cal C}}{d t_b}}\Biggl|_{t_b \rightarrow \infty}= \frac{\phi^{\rm YB} }{G_N^{(2)}lz_h} = \frac{8 \pi S_{\rm YBJT} T_H}{ l} ,
\end{eqnarray}
where $S_{\rm YBJT}=\frac{\phi^{\rm YB}}{4 G_N^{(2)}}$\footnote{This is equivalent to (\ref{Thermal-Entropy}) if we use: $\frac{1}{G_N^{(2)}}=\frac{\Sigma}{G_N^{(3)}}=\frac{\int_{0}^{\pi} d \theta}{G_N^{(3)}}={\rm Limit}_{\theta_{1,2} \rightarrow 0}\left(\frac{\int_{\theta_1}^{\pi -\theta_2} d \theta}{G_N^{(3)}}\right)$. Hence, we can obtain the usual holography from the wedge holography when $\theta_{1,2} \rightarrow 0$.}. Therefore, we see that complexity growth in the presence of Yang-Baxter deformation is proportional to product of the the Bekenstein-Hawking entropy and Hawking temperature of two dimensional black hole.

\section{Conclusion}
\label{summary}
In this paper, we have constructed the wedge holography in homogeneous Yang-Baxter deformed $AdS_3$ background. The same has been done by considering the two Karch-Randall branes located at $\theta=\theta_{1,2}$ in the homogeneous Yang-Baxter deformed bulk $AdS_3$ metric that satisfies Neumann boundary condition on the branes. Let us summarise the key results of the paper.
\begin{itemize}
\item In this setup, the non-conformal boundary of the homogeneous Yang-Baxter deformed $AdS_3$ bulk is located at the holographic screen placed at $z=z_B$. In homogeneous Yang-Baxter deformed wedge holography, two-dimensional field theory is dual to three-dimensional bulk located at the holographic screen which we termed as ``Holographic Screen Non-Conformal Field Theory (HSNCFT)''. Since we have homogeneous Yang-Baxter $AdS_2$ on the Karch-Randall branes and hence the deformed SYK model living at the defect is also situated at the holographic screen because of ``$AdS_2^\eta$/deformed SYK'' duality. Overall, deformed SYK model living at the defect is dual to three-dimensional bulk.

\item According to the wedge holography, the only possible extremal surface is the black hole horizon \cite{GB-3}. The authors in \cite{Massless-Gravity,Massless-Gravity-1,Massless-Gravity-2} obtained non-trivial island surface with entanglement entropy lower than the thermal entropy of the black hole by including the DGP term on the Karch-Randall branes. This possibility has been ruled out by the swampland criteria discussed in \cite{DGP-Swampland, DGP-Swampland-i}. We used the homogeneous Yang-Baxter deformed wedge holography to obtain the Page curve of the black hole and obtained the usual Page curve without DGP term in the presence of homogeneous Yang-Baxter deformation. The homogeneous Yang-Baxter deformation is the reason for the existence of {\it non-trivial island surface} inside the black hole horizon.

\item We also computed the holographic complexity of homogeneous Yang-Baxter deformed $AdS_2$ background using the complexity equals volume proposal \cite{CV}, and find that complexity growth at late times is proportional to the product of Bekenstein-Hawking entropy and temperature of black hole in Yang-Baxter deformed $AdS_2$ background. Without Yang-Baxter deformation, our results reduce to ones obtained in \cite{HC-JT}.

\end{itemize}
In this paper, we restrict ourselves to the homogeneous Yang-Baxter deformed background. Future direction could be the construction of wedge holography and study of holographic complexity in inhomogeneous Yang-Baxter deformed background \cite{Kyono:2017jtc}.

\section*{Acknowledgements}
{
GY is supported by a Senior Research Fellowship (SRF) from the Council of Scientific and Industrial Research (CSIR), Govt. of India. HR would like to thank the authorities of Indian Institute of Technology Roorkee for their unconditional support towards researches in basic sciences. We are grateful to Dibakar Roychowdhury for the fruitful comments and discussions. We would like to thank Rong-Xin Miao, Herman Verlinde and Andreas Karch for the helpful clarifications. 
GY thanks the organizers of ``{\it Holography@25}'' for organizing such a nice event where some part of this work has been done. GY is very grateful to the Science and Engineering Research Board (SERB), Govt. of India (and StAC IIT Roorkee) for providing the partial financial support under the scheme ``{\it International Travel Scheme (ITS)}'' to attend ``{\it Holography@25}''. GY thanks the Infosys foundation for the partial support at CMI.
}
\appendix
\section{Hawking temperature of black hole in homogeneous Yang-Baxter deformed $AdS_2$ background}
\label{Thermodynamics-YBJT}
Black hole solution is given as:
\begin{eqnarray}
\label{metric-BH-h}
ds^2_\eta=-g_{tt}(z)dt^2+g_{zz}(z)dz^2={\cal F}_\eta(z)\Biggl(\frac{-f(z)dt^2+\frac{dz^2}{f(z)}}{z^2 }\Biggr),
\end{eqnarray}
where ${\cal F}_\eta(z)=\frac{1}{1-\frac{\kappa^2}{z^2}},\ f(z)=1-\frac{z^2}{z_h^2}\hspace{1mm}$. 
We can calculate the Hawking temperature using the following formula \cite{Yadav:2022qcl}:
\begin{eqnarray}
\label{Hawking-temp-formula}
T_H=\frac{\kappa}{2 \pi}=\Biggl|Limit_{z \rightarrow z_h}\frac{\partial_z g_{tt}(z)}{4 \pi \sqrt{g_{tt}(z)g_{zz}(z)}}\Biggr|.
\end{eqnarray}
Using (\ref{metric-BH-h}), we obtained the Hawking temperature after simplification of (\ref{Hawking-temp-formula}) as follows:
\begin{eqnarray}
\label{Hawking-temp-YBJT}
T_H=\frac{1}{2 \pi z_h}.
\end{eqnarray}
%%%%%%%%%
%%%%%%%%%%%%%%%
\section{Dimensional reduction from $3D$ to $2D$} 
\label{3D-to-2D}
Three dimensional action is given by \cite{YB-AdS3}\footnote{Terms on Karch-Randall branes are trivial in bulk action (\ref{action-YBD}) because ${\cal K}_\gamma=T_\gamma=0$.}:
\begin{eqnarray}
\label{3D}
& & S_{\rm 3D} =\frac{1}{16 \pi G_N^{(3)}}\Biggl(\int_M d^3x \sqrt{-g^{(3)}} (R^{(3)}-2 \Lambda) \Biggr),
\end{eqnarray}
Upon dimensional reduction along $\theta$, using the ansatz: $ds^2=g^{(3)}_{MN} dx^M dx^N=h^{(2)}_{ij} dx^i dx^j+ {\cal G}_{\theta \theta}^\eta d\theta^2 \equiv h^{(2)}_{ij} dx^i dx^j+ \left(\phi^{\rm YB}\right)^2 d\theta^2$, three dimensional Ricci scalar is obtained as: $R^{(3)}=R+\phi_\eta$ (where $R$ is two dimensional Ricci scalar), and hence action (\ref{3D}) reduces to:
{\footnotesize
\begin{eqnarray}
\label{2D}
& & \hskip -0.3in S_{\rm 2D} =\frac{\Sigma_I}{16 \pi G_N^{(3)}}\Biggl(\int d^2x \sqrt{-h^{(2)}} \phi^{\rm YB} (R+\phi_\eta-2 \Lambda) \Biggr) \equiv \frac{1}{16 \pi G_N^{(2)}}\Biggl(\int d^2x \sqrt{-h^{(2)}} \phi^{\rm YB} (R+\phi_\eta-2 \Lambda) \Biggr),
\end{eqnarray}
}
where $\Sigma_I$ being the volume of compact space and
\begin{eqnarray}
& & 
\frac{1}{G_N^{(2)}}=\frac{\Sigma_I}{ G_N^{(3)}}, \ \ \phi^\eta=\frac{z^2\left(\left(\partial_t {\cal G}_{\theta \theta}^\eta \right)^2-\left(\partial_z {\cal G}_{\theta \theta}^\eta \right)^2+2 {\cal G}_{\theta \theta}^\eta \left(- \partial_t^2+\partial_z^2 \right)  {\cal G}_{\theta \theta}^\eta \right)}{4 {\cal F}_\eta(z) \left({\cal G}_{\theta \theta}^\eta \right)^2}.
\end{eqnarray}
It has been shown in \cite{YB-AdS3} that $\sqrt{{\cal G}_{\theta \theta}^\eta} \phi_\eta= \phi^{\rm YB} \phi_\eta =- U\left(\Phi_\eta^2\right)$, i.e. one obtains the dilaton potential of \cite{Kyono:2017jtc}. Hence,
\begin{eqnarray}
\label{2D-i}
& &  S_{\rm 2D} =\frac{1}{16 \pi G_N^{(2)}}\Biggl[\int d^2x \sqrt{-h^{(2)}} \Biggl( \phi^{\rm YB} (R-2 \Lambda)- U\left(\Phi_\eta^2\right)\Biggr)\Biggr].
\end{eqnarray}
Variation of (\ref{2D-i}) with respect to $h_{ij}$ leads to the Einstein's equation with negative cosmological constant in two dimensions:
\begin{align}\label{2deom-1}
    R_{ij}-\frac{R[h_{ij}]}{2}h_{ij}+\Lambda h_{ij}=0.
\end{align}

{

}

\begin{thebibliography}{99}
%%%%%%%%%%%%%%%
\bibitem{AdS-CFT} J. M. Maldacena, Int.
J. Theor. Phys. 38, 1113-1133 (1999).
%%%%%%%%%%
\bibitem{Hawking} S.W.~Hawking, Commun. Math. Phys. 43, 199 (1975) Erratum: [Commun. Math. Phys. 46, 206 (1976)].
%%%%%
\bibitem{Hawking1} S.W.~Hawking, Phys. Rev. D {\bf 14} (1976) 2460-2473.
%%%%%%%%%%
\bibitem{Page} D.N.~Page, Phys.Rev.Lett. 71 (1993) 3743-3746.
%%%%%%%%%%%%%%%%%
\bibitem{AMMZ} A.~Almheiri, R.~Mahajan, J.~Maldacena and Y.~Zhao, JHEP {\bf 03} (2020) 143.
%%%%%%%%%%%%%%%
\bibitem{Swansea} T.~J.~Hollowood, S.~P.~Kumar, J. High Energ. Phys. 2020, 94 (2020); T.J. Hollowood, S.~P.~Kumar and A. Legramandi, J.Phys.A 53 (2020) 47, 475401; T.J. Hollowood, S.~P.~Kumar, A. Legramandi and N.~Talwar, J. High Energ. Phys. 2021, 67 (2021); T.J. Hollowood, S.~P.~Kumar, A. Legramandi and N.~Talwar, J. High Energ. Phys. 2022, 78 (2022); T.J. Hollowood, S.~P.~Kumar, A. Legramandi and N.~Talwar, J. High Energ. Phys. 2022, 110 (2022); Z. Gyongyosi, T.J. Hollowood, S.P. Kumar, A. Legramandi and N. Talwar, J. High Energ. Phys. 2023, 139 (2023); K.~Goswami and K.~Narayan, JHEP {\bf 10} (2022) 031; F.~Omidi, JHEP \textbf{04}, 022 (2022).
%%%%%%%%%%
\bibitem{RNBH-HD} G.~Yadav, Eur. Phys. J. C {\bf 82}  (2022) 904.
%%%%%%%%%%%
\bibitem{Gopal+Nitin} G.~Yadav and N.~Joshi, Phys. Rev. D {\bf 107}, 026009 (2023).

%%%%%%%%%%%
\bibitem{HD-Page Curve-2} G.~Yadav and A.~Misra,  Phys. Rev. D {\bf 107}, 106015 (2023).
%%%%%%%%%%%%
%\cite{Chou:2021boq}
\bibitem{Chou:2021boq}
C.~J.~Chou, H.~B.~Lao and Y.~Yang, Phys. Rev. D \textbf{106}, no.6, 066008 (2022).
%%%%%%%%% 
%\cite{Chou:2023adi}
\bibitem{Chou:2023adi}
C.~J.~Chou, H.~B.~Lao and Y.~Yang, arXiv:2306.16744 [hep-th].
%%%%%%%
\bibitem{WH-ii} P.~J.~Hu and R.~X.~Miao, JHEP {\bf 03} (2022)145.
%%%%%%%%%%
\bibitem{WH-iii} H.~Geng, A.~Karch, C.~P.~Pardavila, S.~Raju, L.~Randall, M.~Riojas, and S.~Shashi, Phys.Rev.Lett. 129 (2022) 23, 231601.
%%%%%
\bibitem{WH-i} N.~Ogawa, T.~Takayanagi, T.~Tsuda and T.~Waki, arXiv:2207.06735.
%%%%%%%%%%%%%
\bibitem{Hartman-Maldacena} T.~Hartman and J.~Maldacena, JHEP {\bf 05} (2013) 014.
%%%%%%%%
\bibitem{massive-gravity} O.~Aharony, O.~DeWolfe, D.Z.~Freedman and A.~Karch, JHEP {\bf 07} (2003) 030.
%%%%%%%
\bibitem{GB-2} H.~Geng and A.~Karch, JHEP {\bf 09} (2020) 121.
%%%%%%%%%%%%%%%%%%%%%%%%%%%%
\bibitem{GB-3} H.~Geng, A.~Karch, C.~P.~Pardavila, S.~Raju and L.~Randall, SciPost Phys. 10 (2021) 5, 103.
%%%%%%%
\bibitem{GB-4} H.~Geng, A.~Karch, C.~P.~Pardavila, S.~Raju and L.~Randall, JHEP 01 (2022) 182.
%%%%%%%%%%%%%%
\bibitem{Massless-Gravity} R.~X.~Miao, arXiv:2212.07645[hep-th].
%%%%%%%%%%%%%
\bibitem{C1} K.~Ghosh and C.~Krishnan, JHEP 08 (2021) 119.
%%%%%%%%%%%
\bibitem{Massless-Gravity-1} R.~X.~Miao, JHEP {\bf 03} (2023) 214.
%%%%%%%%%%%
\bibitem{Massless-Gravity-2} C.~P.~Pardavila,  arXiv:2302.04279[hep-th].
%%%%%%%%
\bibitem{Massless-Gravity-3} D.~Li and R.~X.~Miao, arXiv:2303.10958[hep-th].
%%%%%%%%%%%%%%%
 \bibitem{DGP-2} G.~R.~Dvali, G.~Gabadadze and M.~Porrati, Phys. Lett. B 485 (2000) 208.
 %%%%
\bibitem{Swampland} C.~Vafa, arXiv:hep-th/0509212; N.~Arkani-Hamed, L.~Motl, A.~Nicolis and C.~Vafa, JHEP {\bf 06} (2007) 060; H.~Ooguri and C.~Vafa, Nucl.
Phys. B {\bf 766} (2007) 21.
%%%%%%%%%
\bibitem{DGP-Swampland} H.~Geng, A.~Karch, C.~P.~Pardavila, L.~Randall, M.~Riojas, S.~Shashi and M.~Youssef, arXiv:2306.15672.
%%%%%%
\bibitem{DGP-Swampland-i} H.~Geng, arXiv:2306.15671.
%%%%%%%
\bibitem{Yadav:2020tyo} A.~Misra and V.~Yadav, arXiv:2004.07259 [hep-th].
%%%%%%%%%%%%%%%%%%%
\bibitem{Multiverse} G.~Yadav, JHEP {\bf 03} (2023) 103.
%%%%
 \bibitem{Matsumoto:2015jja}
T.~Matsumoto and K.~Yoshida, Nucl. Phys. B \textbf{893}, 287-304 (2015);  C.~Klimcik,
JHEP \textbf{12}, 051 (2002);
C.~Klimcik,  J. Math. Phys. \textbf{50}, 043508 (2009);
F.~Delduc, M.~Magro and B.~Vicedo,
JHEP \textbf{11}, 192 (2013);
J.~Pal, H.~Rathi, A.~Lala and D.~Roychowdhury, arXiv:2208.09599 [hep-th].
%%%%%%
\bibitem{DR-1} D.~Roychowdhury, JHEP {\bf 10} (2017) 056; D.~Roychowdhury, Phys. Lett. B {\bf 778} (2018) 167–173; A.~Banerjee, A.~Bhattacharyya and D.~Roychowdhury, JHEP {\bf 02} (2018) 035; D.~Roychowdhury, JHEP {\bf 05} (2018) 018.
%%%%%%%%%%%%
%%%%%%%
\bibitem{Kyono:2017jtc}
H.~Kyono, S.~Okumura and K.~Yoshida, JHEP \textbf{03}, 173 (2017);
H.~Kyono, S.~Okumura and K.~Yoshida, Nucl. Phys. B \textbf{923}, 126-143 (2017);
S.~Okumura and K.~Yoshida, Nucl. Phys. B \textbf{933}, 234-247 (2018).
%%%
\bibitem{Almheiri:2014cka}
A.~Almheiri and J.~Polchinski, JHEP \textbf{11}, 014 (2015).
%%%%
\bibitem{YB-AdS3} A.~Lala and D.~Roychowdhury, JHEP {\bf 12} (2018) 073.
%%%%%%%%%%%%%%
\bibitem{Wald-i}R. M.~Wald, Phys. Rev. D {\bf 48}, 3427 (1993).
%%%%%%%%%%%%%%%%%%%
\bibitem{Wald-ii}V.~Iyer and R. M.~Wald, Phys. Rev. D {\bf 50}, 846 (1994).
 %%%%%%%%%%%%
\bibitem{Miao:2020oey}
R.~X.~Miao, JHEP \textbf{01}, 150 (2021).
%%%%%%%%%
\bibitem{Akal:2020wfl}
I.~Akal, Y.~Kusuki, T.~Takayanagi and Z.~Wei, Phys. Rev. D \textbf{102}, no.12, 126007 (2020).
 %%%%%%%%%%
 \bibitem{Roychowdhury:2017oqd}
D.~Roychowdhury, JHEP \textbf{03}, 043 (2017).
%%%%%%%%%%%%%%%%
 \bibitem{HS-Chethan} C.~Krishnan, V.~Patil and J.~Pereira, arXiv:2005.02993.
%%%%%%%%%%%%%%%%%
\bibitem{KR1} A.~Karch and L.~Randall, JHEP {\bf 05} (2001) 008.
%%%%%%%%%%%%%
\bibitem{KR2} A.~Karch and L.~Randall, JHEP {\bf 06} (2001) 063.
%%%%%%%%%
\bibitem{Sachdev:1992fk}
S.~Sachdev and J.~Ye, Phys. Rev. Lett. \textbf{70}, 3339 (1993) [arXiv:cond-mat/9212030 [cond-mat]]; A.Kitaev.2015. A simple model of quantum holography, talk given at KITP strings seminar and Entanglement program, February 12, April 7, and May 27, Santa Barbara, U.S.A.;
J.~Maldacena and D.~Stanford, Phys. Rev. D \textbf{94}, no.10, 106002 (2016);
J.~Polchinski and V.~Rosenhaus, JHEP \textbf{04}, 001 (2016);
K.~Jensen, Phys. Rev. Lett. \textbf{117}, no.11, 111601 (2016);
S.~R.~Das, A.~Jevicki and K.~Suzuki, JHEP \textbf{09}, 017 (2017);
S.~R.~Das, A.~Ghosh, A.~Jevicki and K.~Suzuki, JHEP \textbf{02}, 162 (2018);
D.~Roychowdhury, Phys. Lett. B \textbf{797}, 134818 (2019);
J.~Maldacena, D.~Stanford and Z.~Yang, PTEP \textbf{2016}, no.12, 12C104 (2016);
A.~Goel, V.~Narovlansky and H.~Verlinde, [arXiv:2301.05732 [hep-th]].
%%%%%%%%
\bibitem{NGB} Q.~L.~Hu, D.~Li, R.~X.~Miao and Y.~Q.~Zhao, JHEP {\bf 09} (2022) 037.
%%%%%%%%%%%%%%%%%%%%%%%%%%%%%%%%%%%%%%%%%%%%%%%
\bibitem{RT} S.~Ryu, T.~Takayanagi, Phys. Rev. Lett. {\bf 96}, 181602 (2006).
%%%%%%%%%
\bibitem{I-3} W.~C.~Gan,  D.~H.~Du and F.~W.~Shu, JHEP {\bf 07} (2022) 020.
%%%%% 
\bibitem{scrambling-time-1} P.~Hayden and J.~Preskill, JHEP, 0709:120,2007.
%%%%%%%%%%%%%%
\bibitem{scrambling-time-2} Y.~Sekino and L.~Susskind, JHEP 0810:065,2008.
%%%%%%%%%%%%
\bibitem{scrambling-time-EW} G.~Penington, arXiv:1905.08255[hep-th].
%%%%%%%%%%%%%{}
\bibitem{Ankit} A.~Anand, arXiv:2205.13785[hep-th].
%%%%%%%%%
\bibitem{CV} D.~Stanford and L.~Susskind, Phys. Rev. D 90, 126007 (2014).
%%%%%%%%%
\bibitem{HC-JT} A.~Bhattacharya, A.~Bhattacharyya and A.~K.~Patra, arXiv:2304.09909[hep-th].
%%%%%%%%
%\bibitem{Kyono:2017jtc}
%H.~Kyono, S.~Okumura and K.~Yoshida, {\it Deformations of the Almheiri-Polchinski model}, JHEP \textbf{03}, 173 (2017) [arXiv:1701.06340 [hep-th]].
%%%%%%%%%%
\bibitem{Yadav:2022qcl}
G.~Yadav, Phys. Lett. B \textbf{841}, 137925 (2023).
%%%%%%
%\bibitem{Almheiri:2014cka}
%A.~Almheiri and J.~Polchinski, {\it Models of AdS$_{2}$ backreaction and holography}, JHEP \textbf{11}, 014 (2015) [arXiv:1402.6334 [hep-th]].
\end{thebibliography}
\end{document}